# Nonlinear Breakdown of Antisymmetric Flexoelectric Coupling

Yingzhuo Lun[1, *], Yida Yang[1], Tingjun Wang[1], Qi Ren[1], Sang-Wook Cheong[2], Xueyun Wang[1], and Jiawang Hong[1, *]

[1]School of Aerospace Engineering, Beijing Institute of Technology, Beijing 100081, China

[2]Rutgers Centre for Emergent Materials and Department of Physics and Astronomy, Rutgers University, New Jersey 08854, USA

[*]Correspondence to: lunyingzhuo@bit.edu.cn (Y.L.), hongjw@bit.edu.cn (J.H.)

**Highlight:**

Nonlinear coupling reveals an intrinsic symmetry-breaking flexoelectric response beyond the conventional antisymmetric framework.

**Abstract**

Flexoelectric coupling is conventionally regarded as antisymmetric, such that reversing the strain gradient reverses the polarization direction while preserving its magnitude. Here we report that this antisymmetric coupling breaks down in noncentrosymmetric single crystals driven by large strain gradients, where nonlinearity becomes operative. Harmonic-resolved measurements reveal robust even-order flexoelectric harmonics that are forbidden in antisymmetric systems, as well as the three-wave mixing indicative of symmetry breaking under dual-frequency excitation. These frequency-relevant generations provide evidence for asymmetric flexoelectricity in the nonlinear regime. A reconstructed nonlinear constitutive framework consistently accounts for all observations by incorporating even-order strain-gradient terms. Our findings establish asymmetric flexoelectricity as an intrinsic electromechanical response and uncover a symmetry-breaking mechanism in flexoelectricity.

## Introduction

Flexoelectricity describes the electromechanical coupling between electric polarization and strain gradient [1,2] and is universally allowed in all insulating materials, independent of crystal symmetry [3]. Unlike piezoelectricity, which requires broken inversion symmetry, flexoelectricity naturally emerges from spatially nonuniform deformation and becomes increasingly prominent at reduced length scales [4,5] where large strain gradients can be readily generated. This unique feature has enabled a wide range of strain-gradient-driven functionalities across various material systems, including centrosymmetric dielectrics [6,7], ferroelectrics [8,9], polymers [10,11], and low-dimensional nanostructures [12,13]. Flexoelectric coupling has been implicated in a variety of emerging physical phenomena, such as polarization enhancement [14], mechanical domain switching [15,16], asymmetric mechanical properties [17,18], flexoelectronics [19,20] and flexo-photovoltaic [21,22], *etc*. In these phenomena, the electromechanical response is governed by the constitutive relation between polarization and strain gradient, and the symmetry of this relation under strain-gradient inversion is consequently essential for interpreting the strain-gradient-driven mechanisms.

Conventionally, flexoelectric response has been described as antisymmetric [23–25], such that the induced polarization is an odd function of the strain gradient, a representative example being its linear response. Under this assumption, reversing the strain gradient reverses the polarization direction while preserving its magnitude, thereby enforcing a strictly antisymmetric relation between polarization and strain gradient. This antisymmetric assumption has been adopted and serves as the theoretical foundation for interpreting various flexoelectric phenomena. In this context, asymmetric flexoelectricity—defined as a polarization response that is no longer invariant in magnitude under strain-gradient inversion—is implicitly excluded. Recent studies have begun to experimentally explore nonlinear flexoelectric effects under large strain gradients, revealing higher-order responses beyond the linear regime [26]. However, these nonlinear contributions have largely been treated as higher-order corrections that retain the underlying antisymmetric character. Theoretically, it has been suggested that quadratic nonlinearity may arise in non-centrosymmetric materials [27,28], potentially challenging the antisymmetric assumption, yet direct experimental validation remains absent. Therefore, a fundamental question remains unresolved: Does the antisymmetric assumption of flexoelectricity remain valid in the nonlinear regime, or does nonlinear coupling fundamentally alter the symmetry of the polarization–strain-gradient relation?

In this work, we demonstrate that nonlinear coupling drives a breakdown of the antisymmetric assumption of flexoelectricity in noncentrosymmetric single crystals under large strain gradients. Using harmonic-resolved flexoelectric measurements, we identify frequency-relevant generations that are forbidden in an antisymmetric system, including robust even-order harmonics and three-wave mixing indicative of asymmetry. To interpret these observations, we construct a nonlinear constitutive framework incorporating even-order strain-gradient terms that consistently captures the experimental behavior, and provides a fundamental theoretical description of the asymmetric flexoelectricity. These findings refine the symmetry understanding of flexoelectricity in the nonlinear regime.

**Methods and Results**

To probe the symmetry of flexoelectricity, we design a harmonic-resolved flexoelectric measurement based on three-point bending [29] of a poled ferroelectric single crystal of $0.7Pb(Mg_{1/3}Nb_{2/3})O_3$-$0.3PbTiO_3$ (labelled as PMN-30PT), as illustrated in Fig. 1. Periodic mechanical bending generates a well-defined oscillatory strain gradient across the beam thickness while the excited flexoelectric current response is recorded and subsequently analyzed by fast Fourier transform, simultaneously resolving the fundamental and all high-order harmonic generations. This approach provides a sensitive probe of symmetry: an antisymmetric flexoelectric coupling strictly permits only odd-order harmonics, whereas the emergence of even-order harmonics would unambiguously signal a symmetry breaking.

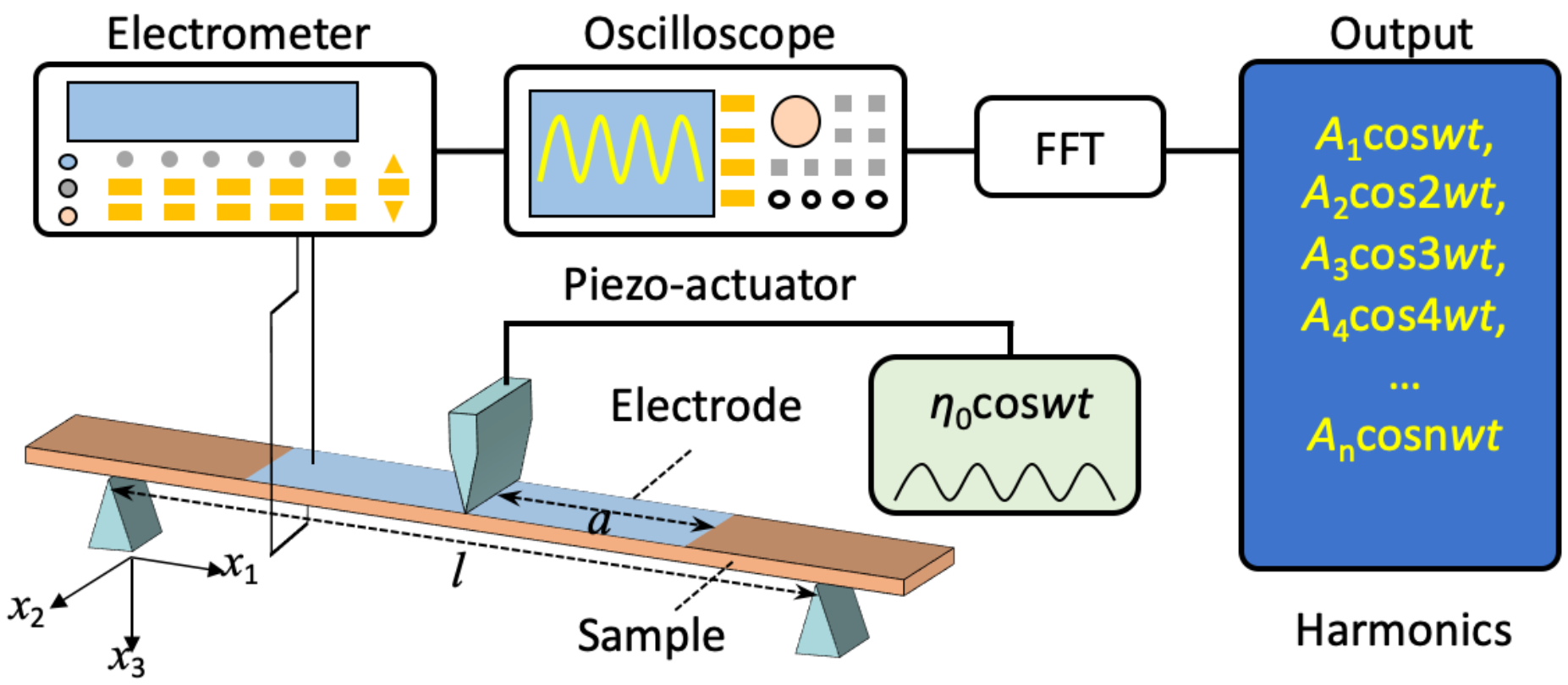

**Figure 1. Experimental setup for harmonic-resolved flexoelectric measurement.** The size of sample is set as 25 mm long, 3 mm wide, and 500 μm thick. Silver electrodes of area 36 $mm^2$ with half-length $a$ = 6 mm are deposited on the top and bottom surfaces for current collection. The distance between the two simple-support bases is set as $l$ = 20 mm. The flexoelectric current wave is measured using electrometer (Keithley 6514) and oscilloscope (Tektronix TBS2000).

We first examine the flexoelectric response under small strain-gradients with a single-frequency excitation at 5 Hz. At a strain gradient amplitude of $\eta_0 = 0.019\ m^{-1}$, the measured flexoelectric current follows a cosinusoidal waveform that faithfully follows the applied bending excitation (Fig. 2a, top panel). The corresponding frequency spectrum contains only a single component at the driving frequency, with no discernible higher-order harmonics (Fig. 2b). This behavior is consistent with the conventional understanding of flexoelectricity as a linear and antisymmetric coupling. In contrast, when the strain gradient is increased by one order of magnitude ($0.19\ m^{-1}$), the current waveform becomes strongly non-cosinusoidal, and its frequency spectrum reveals substantial high-order harmonic components beyond the driving frequency. The high-order harmonics indicates that the flexoelectric response can no longer be described as a purely linear transduction, while the involvement of nonlinear coupling becomes significant. In addition to the expected odd-order harmonics, robust even-order harmonics—most notably the second and fourth—emerge and grow rapidly as the strain gradient further increases to 0.26 m$^{-1}$. The emergence of symmetry-forbidden even-order harmonics indicate a breakdown of antisymmetric flexoelectric coupling beyond the linear regime. The corresponding frequency spectrum of polarization (Fig. 2c) shows an attenuation of harmonic amplitude with increasing order, implying that the signal originates from intrinsic electromechanical transduction rather than electrical noise.

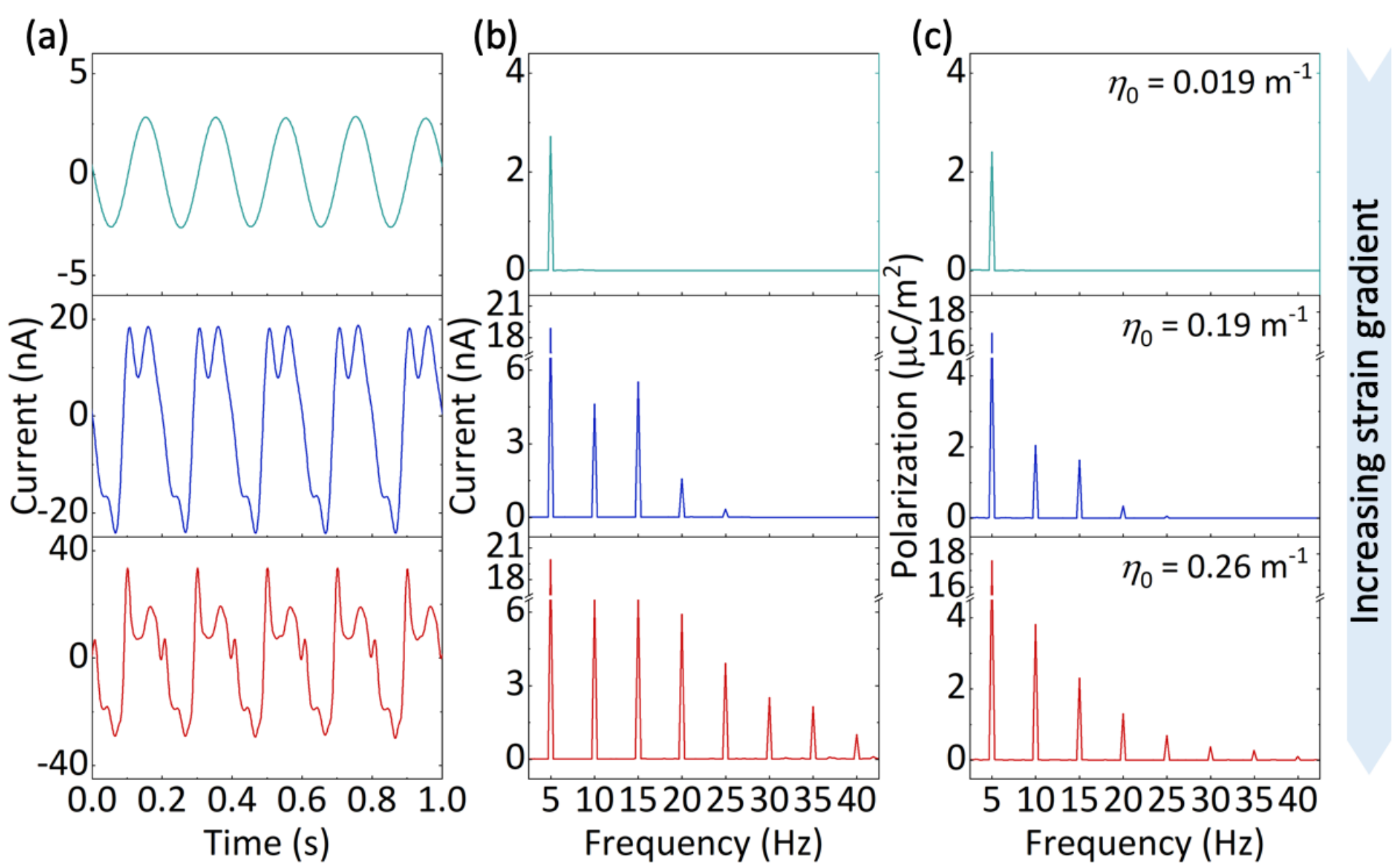

**Figure 2. High-order harmonic generations excited by increasing strain gradients. (a)** Time-dependent flexoelectric current waveform excited by increasing strain gradients. (**b-c**) Corresponding fast Fourier transform analyzed frequency spectrums of flexoelectric current and polarization, respectively.

To demonstrate the relation between higher-order harmonics and nonlinear flexoelectric coupling, we systematically quantify the dependence of each harmonic components on the strain gradient. Figure 3 summarizes the amplitudes of the first four harmonics as a function of strain gradient, with the maximum gradient deliberately limited below 0.19 m$^{-1}$ to ensure that only the low-order nonlinear contributions are appreciably activated for simplification. At small strain gradients, the first harmonic dominates and exhibits an approximately linear dependence (Fig. 3a), consistent with the linear flexoelectricity. As the strain gradient increases, the first harmonic gradually deviates from linearity, while higher-order harmonics become increasingly pronounced (Fig. 3b-d). Notably, the higher-order harmonics exhibits distinct different scaling behaviors (revealed by polynomial fitting): the second (2$^{nd}$), third (3$^{rd}$), and fourth (4$^{th}$) harmonics scale predominantly as the second, third and fourth power of the strain gradient, respectively. This clear hierarchy of power-law dependencies cannot be produced in linear system, and instead constitutes direct evidence of high-order nonlinear flexoelectric coupling. Particularly, the quadratic (quartic) scaling of the second (fourth) harmonic evidences the presence of even-order coupling terms, which introduce polarization components that are invariant under strain-gradient inversion, and thereby breaks symmetry when superimposed with odd-order terms. The rapidly growth of these even-order coupling responses therefore establish that the flexoelectric coupling becomes asymmetric in nonlinear regime.

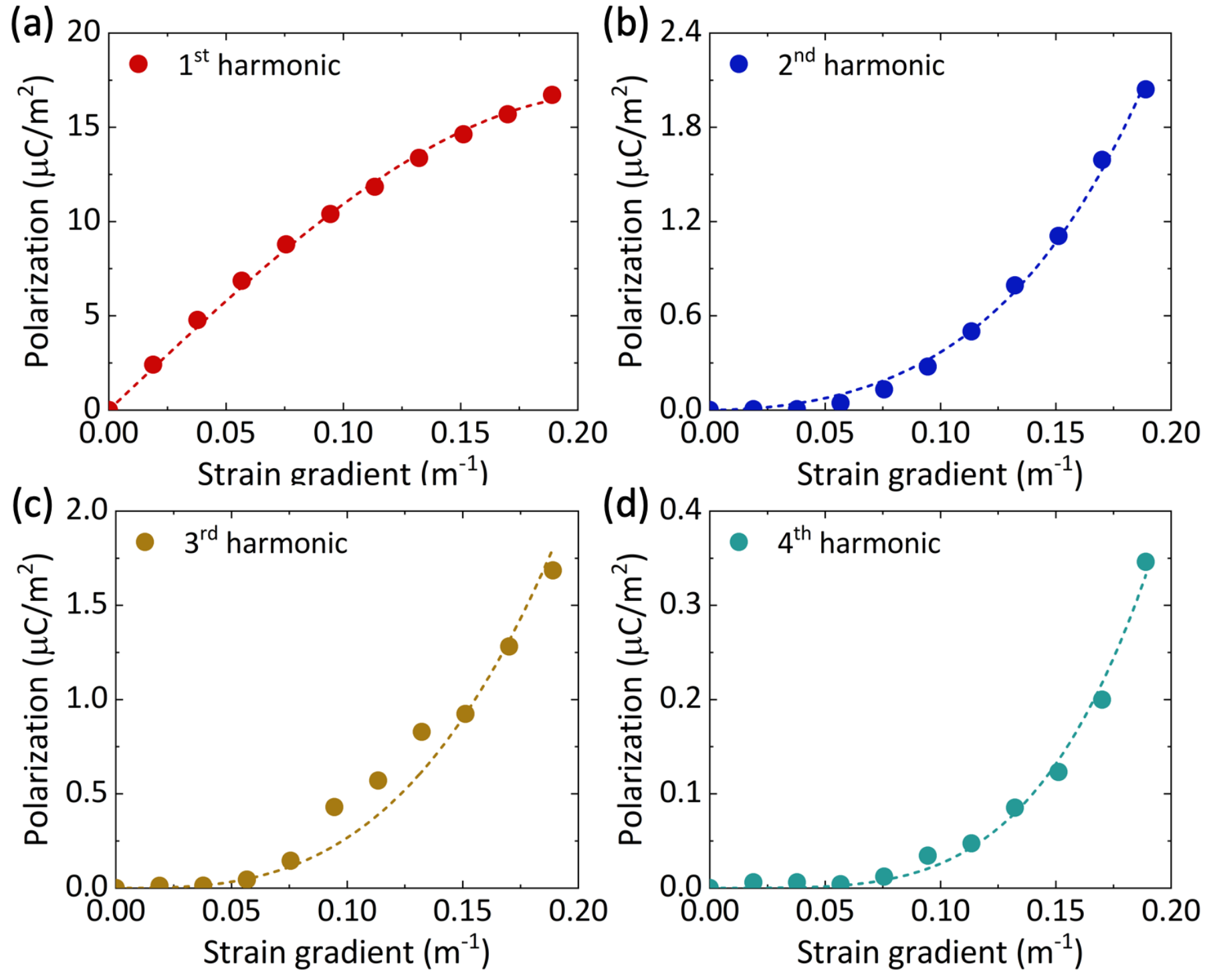

**Figure 3. Nonlinear dependence of flexoelectric harmonics on strain gradient**. (**a**) First, (**b**) second, (**c**) third, and (**d**) fourth harmonic amplitude as a function of strain gradient, simultaneously extracted from the frequency spectrums excited by each strain gradient. The dash lines show the fitting of raw experimental data following (a) linear-cubic, (b) quadratic-quartic, (c) cubic, and (d) quartic polynomials.

To provide more evidences for the existence of even-order nonlinear coupling, we employ dual-frequency excitation which provides a more rigorous verification through frequency conversion. Specifically, even-order coupling terms enable the coherent generation of special sum- and difference-frequency components through nonlinearity-mediated three-wave mixing, which thereby constitutes a direct signature of asymmetric flexoelectricity. Herein, two independent driving frequencies ($\omega_1$ = 5 Hz and $\omega_2$ = 18 Hz) are simultaneously applied with comparable strain-gradient amplitudes, as confirmed in top panel of Fig. 4b. The resulting current spectrum (bottom panel in Fig. 4b) contains clear fundamental components at $\omega_1$ and $\omega_2$, accompanied by higher-order harmonics associated with each frequency (marked by blue dots). Notably, even-order harmonics are again observed for both driving frequencies, consistent with the behavior identified under single-frequency excitation. The output spectrum further reveals a series of additional frequency components (marked by red dots) that cannot be attributed to integer multiples of either $\omega_1$ or $\omega_2$. These components appear at frequencies corresponding to coherent sum- and difference-frequency combinations of the two inputs. Their emergence indicates nonlinear interactions between the two driving frequencies, going beyond independent harmonic generation. Importantly, the observed mixing frequencies exhibit characteristics expected for second-order term-mediated three-wave mixing, in particular, distinct components appear at $(\omega_2 - \omega_1)$ and $(\omega_2 + \omega_1)$. These frequency components are forbidden in an antisymmetric framework and can only arise when flexoelectric coupling itself becomes asymmetric. In addition, other higher-order combinations arising from third-order term are also observed.

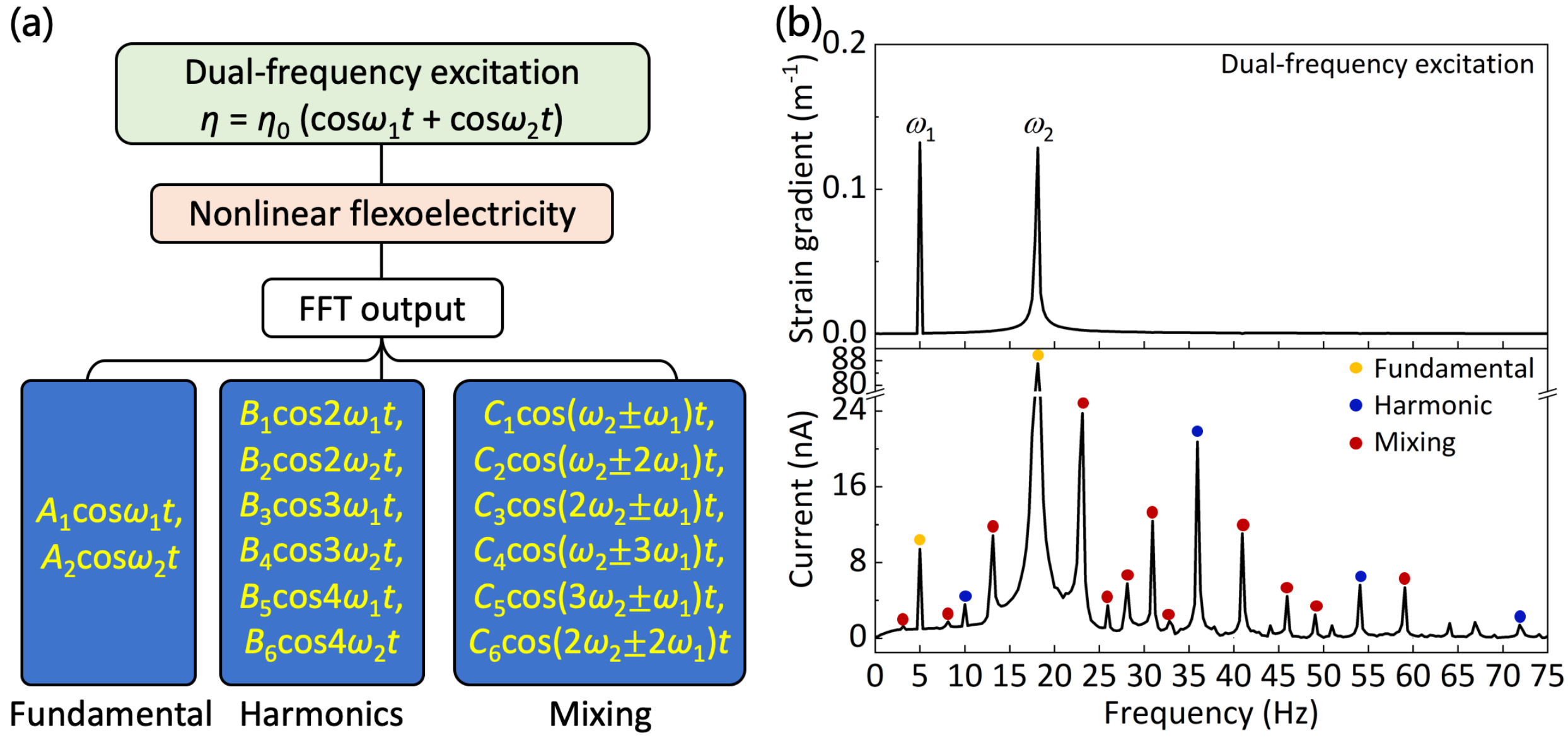

**Figure 4. Even-order nonlinear coupling induced three-wave mixing in dual-frequency excitation.** (**a**) Frequency components allowed in three-wave mixing as predicted from a nonlinear constitutive framework incorporating even-order terms. (**b**) Frequency spectrum of applied strain gradient from the dual-frequency excitation (top panel), and resulting current spectrum exhibits a rich set of frequency components beyond the two driving frequencies (bottom panel).

Finally, we address a critical issue: whether the observed even-order harmonics could originate from other mechanisms, such as ferroelectric or piezoelectric nonlinearity. To this end, identical measurements are performed on unpoled PMN–30PT samples [30], which are effectively centrosymmetric. As shown in Fig. S1, the second- and fourth-order harmonics are strongly suppressed in the unpoled state while the odd-order harmonics remain observable, indicating that the asymmetric flexoelectric response nearly vanishes when macroscopic inversion symmetry is effectively restored by domain averaging. Moreover, the maximum surface strain in response to the strain gradient of $0.26\ m^{-1}$ remains below 0.0065%, far too small to induce nonlinear piezoelectricity and interfere with flexoelectric harmonics, as independently verified by piezoelectric harmonic measurements (Fig. S2). These controls rule out nonlinear piezoelectricity as the dominant origin. Furthermore, we directly examine the asymmetric flexoelectric response by reversing specimen, thereby applying strain gradients of identical magnitude but opposite directions, and different flexoelectric responses are observed. It should be noted, however, that this inversion-based approach is less reliable than harmonic-based methods, as reversing specimen inevitably alters the pre-bending state and the position of loading probe, leading to unequal flexoelectric currents.

Our experimental observations of even-order harmonics and mixing frequency reveal a violation of the conventional antisymmetric framework of flexoelectricity. In the following, we reconstruct a nonlinear constitutive relation that explicitly accounts from symmetry breaking beyond linear regime, and provides a unified explanation for the observed harmonics generation and three-wave mixing. We expand the flexoelectric polarization $P$ as a Taylor series with respect to strain gradient $\eta$,

$$P = \mu_1 \eta + \frac{\mu_2}{2!}\eta^2 + \frac{\mu_3}{3!}\eta^3 + \cdots + \frac{\mu_n}{n!}\eta^n, \tag{1}$$

where $\mu_1$ denotes the linear flexoelectric coefficient, while $\mu_n$ ($n = 2, 4, 6, \ldots$) and $\mu_n$ ($n = 3, 5, 7, \ldots$) describe the even-order and odd-order nonlinear coupling, respectively. Odd-order terms preserve antisymmetry polarization with respect to strain-gradient inversion, whereas even-order terms allow an invariant polarization under strain-gradient inversion. The superposition of two leads to the asymmetric flexoelectricity, *i.e.* $P(+\eta) \neq -P(-\eta)$.

On this basis, we first analyze the consequences of this asymmetric constitutive relation under single-frequency excitation, $\eta = \eta_0 \cos\omega t$ (where $t$ is the time, $\eta_0$ is the average strain gradient, $w$ is the angular frequency). Substituting strain gradient expression into the nonlinear relation Eq. (1) and applying trigonometric identities yields a polarization response containing multiple frequency components (only the first four order terms is considered for simplification),

$$\begin{aligned} P = &\left( \mu_1 \eta_0 + \frac{\mu_3}{8}\eta_0^3 \right)\cos wt + \left( \frac{\mu_2}{4}\eta_0^2 + \frac{\mu_4}{48}\eta_0^4 \right)\cos 2wt \\ &+ \left( \frac{\mu_3}{24}\eta_0^3 \right)\cos 3wt + \left( \frac{\mu_4}{192}\eta_0^4 \right)\cos 4wt + \left( \frac{\mu_2}{4}\eta_0^2 + \frac{\mu_4}{64}\eta_0^4 \right). \end{aligned} \tag{2}$$

The derived expression establishes a relation between the nonlinear terms and the high-order harmonics: the odd-order terms contribute exclusively to odd harmonics, while the even-order terms generate additional components at even multiples of driving frequency. The appearance of second- and fourth-order harmonics thus directly reflects the presence of even-order nonlinear coupling, supporting the observed harmonic spectrum in Fig. 2. Moreover, the amplitude of each harmonic predominantly follows a power-law scaling according to its corresponding nonlinear order. These predictions are in quantitative agreement with experimental measurements (Fig. 3), and thereby enable extraction of flexoelectric coefficients following the Eq. (2), as listed in Table I. The third-order coefficient fitted from the first and third harmonics respectively show a small difference below 8 %, suggesting the measured coefficients are reliable. The reconstructed time-dependent flexoelectric current using the fitted coefficients (Fig. S4) is consistent with the measured wave when $\mu_2$ shows a negative value.

We further extend the analysis to dual-frequency excitation $\eta = \eta_0 (\cos\omega_1 t + \cos\omega_2 t)$, yielding a polarization response Eq. (3) containing mixing frequency components. The derived polarization demonstrates that three-wave mixing arises from the nonlinear terms and the coherent frequencies are uniquely determined by the orders of nonlinearity. In particular, even-order coupling generates polarization components at the sum- and difference-frequencies, i.e., $(\omega_2 + \omega_1)$ and $(\omega_2 - \omega_1)$. Such frequency conversion is strictly forbidden in an antisymmetric constitutive framework (without even-terms) and can only arise from nonlinear terms that break inversion symmetry. The experimentally observed frequency spectrum is consistent with the prediction (as summarized in Fig. 4a), thereby validating the even-order flexoelectric coupling.

$$P = \frac{1}{64}\left[\begin{pmatrix} 32\mu_1\eta_0 \\ +3\mu_3\eta_0^3 \end{pmatrix}\begin{pmatrix} \cos w_1 t \\ +\cos w_2 t \end{pmatrix}\right] + \frac{1}{384}\left[\begin{matrix} \begin{pmatrix} 24\mu_2\eta_0^2 \\ +2\mu_4\eta_0^4 \end{pmatrix}\begin{pmatrix} \cos 2w_1 t \\ +\cos 2w_2 t \end{pmatrix} \\ +\begin{pmatrix} 48\mu_2\eta_0^2 \\ +3\mu_4\eta_0^4 \end{pmatrix}\begin{pmatrix} \cos(w_2 - w_1)t \\ +\cos(w_2 + w_1)t \end{pmatrix} \end{matrix}\right] + \frac{\mu_3}{192}\left[\begin{matrix} \eta_0^3\begin{pmatrix} \cos 3w_1 t \\ +\cos 3w_2 t \end{pmatrix} \\ +3\eta_0^3\begin{pmatrix} \cos(w_2 - 2w_1)t \\ +\cos(w_2 + 2w_1)t \\ +\cos(2w_2 - w_1)t \\ +\cos(2w_2 + w_1)t \end{pmatrix} \end{matrix}\right] + \frac{\mu_4}{3072}\left[\begin{matrix} \eta_0^4\begin{pmatrix} \cos 4w_1 t \\ +\cos 4w_2 t \end{pmatrix} \\ +4\eta_0^4\begin{pmatrix} \cos(w_2 - 3w_1)t \\ +\cos(w_2 + 3w_1)t \\ +\cos(3w_2 - w_1)t \\ +\cos(3w_2 + w_1)t \end{pmatrix} \\ +6\eta_0^4\begin{pmatrix} \cos(2w_2 - 2w_1)t \\ +\cos(2w_2 + 2w_1)t \end{pmatrix} \end{matrix}\right] + \left[\begin{matrix} \frac{\mu_2}{8}\eta_0^2 \\ +\frac{3\mu_4}{512}\eta_0^4 \end{matrix}\right]. \tag{3}$$

**Table I.** Linear and nonlinear flexoelectric coefficients fitted from different harmonics generations

| Coefficients | First | Second | Third | Fourth |
|---|---|---|---|---|
| $\mu_1$ (μC·m$^{-1}$) | 118 | | | |
| $\mu_2$ (μC) | | -114 | | |
| $\mu_3$ (μC·m) | $-6.9\times10^3$ | | $-6.4\times10^3$ | |
| $\mu_4$ (μC·m$^2$) | | $-4.1\times10^4$ | | $5.0\times10^4$ |

**Discussions**

The established nonlinear constitutive framework demonstrates the manifestation of asymmetric flexoelectricity is subject to well-defined symmetry and conditions. At small strain gradients, the flexoelectric response remains dominated by the linear term, preserving the conventional antisymmetric behavior long assumed in flexoelectric studies. As the strain gradient increases, nonlinear coupling becomes significant and activate symmetry-breaking responses. Importantly, even-order nonlinear coupling contributes terms that are associated with odd-rank tensors, and symmetry arguments dictate that all odd-rank tensor components are strictly forbidden in materials with inversion symmetry [31,32]. As a result, asymmetric flexoelectricity is expected to only occur in non-centrosymmetric systems, consistent with our experiments. This symmetry limitation distinguishes the present phenomenon from general nonlinear effects and implies its intrinsic origin. From a broader perspective, the identification of asymmetric flexoelectricity highlights a general physical principle: nonlinear coupling can modify not only the magnitude but also the symmetry of a response function. Similar symmetry-evolving behavior is well-established in nonlinear optical systems [33,34], where higher-order interactions open up responses forbidden in the linear regime. Extending this perspective to flexoelectricity suggests that symmetry-breaking enable by nonlinearity may play an unrecognized role in a broad class of strain-gradient–mediated phenomena.

## Conclusions

In summary, we demonstrate that the long-assumed antisymmetric form of flexoelectric coupling is no longer preserved in nonlinear regime driven by large strain gradient. The observations of robust even-order flexoelectric harmonics and symmetry-forbidden primary three-wave mixing establish direct evidences for the emergence of asymmetric flexoelectric coupling. By developing a nonlinear constitutive framework that explicitly incorporates even-order strain-gradient terms, we identify asymmetric flexoelectricity as an intrinsic electromechanical response. These findings redefine the symmetry framework of flexoelectricity, and highlight nonlinearity-induced symmetry-breaking as a fundamental aspect of strain-gradient-mediated electromechanical interactions.

## Declaration of Competing Interest

The authors declare that they have no competing interests.

## Acknowledgments

This work is supported by the National Natural Science Foundation of China (Grant Nos. 12402183, 12404101), Beijing Natural Science Foundation (Grant No. 1244057), and National Key Projects for Research and Development of China (Grant No. 2021YFA1400300).

## Data availability

There are no publicly available research data or software supporting this manuscript. Requests for further information or data should be sent to the authors.